\documentclass[pra,aps,10pt,superscriptaddress,twocolumn]{revtex4-2}
\usepackage[colorlinks=true, linkcolor=blue, citecolor=blue, urlcolor=blue]{hyperref}
\usepackage{graphicx,color}
\usepackage{amsmath,amssymb,bm,bbm}
\usepackage{amsmath}
\usepackage{subfigure}

\newcommand{\Hq}{\mathcal{H}}

\newcommand{\hs}{\mathcal{H}_{\text{\tiny S}}}
\newcommand{\hl}{\mathcal{H}_{\text{\tiny L}}}
\newcommand{\hsl}{\mathcal{H}_{\text{\tiny SL}}}

\newcommand{\nn}{\nonumber}

\newcommand{\tr}{\text{Tr}}
\newcommand{\rh}{\rho_{s}}

\newcommand{\etal}{\emph{et al.}}

\begin{document}

\title{Emergence of Unruh prethermalization for uniformly accelerating many-atom system}

\author{Saptarshi Saha}
\email{s.saha@tu-berlin.de}
\affiliation{Department of Physical Sciences, Indian Institute of Science
Education and Research Kolkata, Mohanpur - 741 246, WB, India}
\affiliation{Institut für Physik und Astronomie, Technische Universität Berlin, Hardenbergstr. 36, 10623 Berlin, Germany}

\author{Chiranjeeb Singha}
\email{chiranjeeb.singha@iucaa.in}
\affiliation{Inter-University Centre for Astronomy \& Astrophysics, Post Bag 4, Pune 411 007, India}

\author{Pragna Das}
\email{Pragna.Das@ijs.si}
\affiliation{Indian Institute of Science Education and Research Bhopal 462066 India}
\affiliation{Department of Theoretical Physics, J. Stefan Institute, SI-1000 Ljubljana, Slovenia}

\author{Arpan Chatterjee}
\email{arpanchatterjee@ufscar.br}
\affiliation{Department of Physical Sciences, 
Indian Institute of Science Education and Research Kolkata,
Mohanpur - 741 246, WB, India}
\affiliation{Departamento de Física, Universidade Federal de São Carlos, Rodovia Washington Luís,
km 235—SP-310, 13565-905 São Carlos, SP, Brazil}

\date{\today}

\begin{abstract}
A uniformly accelerated atom in an inertial vacuum generally thermalizes and reaches a Gibbs state. This phenomenon is commonly known as the Unruh effect. Here, we show that the situation is entirely different for the many-atoms problem. In the case of non-interacting accelerating atoms, we show that a regime exists where the entire system reaches a prethermal generalized Gibbs state before it thermalizes. The prethermal state is protected by emergent conserved quantities; hence, the system behaves like a nearly-integrable one, which shows a sharp distinction from the Unruh effect. We coin the term ``Unruh prethermalization" to characterize this phenomenon. The measure of entanglement is a good estimation of the lifetime of the prethermal state and is consistent with previous studies. Finally, we show that in such a regime, the dynamics show a Dicke superradiance-type radiation burst before reaching the prethermal state. In contrast, only a mono-exponential decay is observed for Unruh thermalization. In addition, to highlight the significance of our results, we compare them with existing experimental observations.
\end{abstract}

%
\maketitle

\section{Introduction}  Thermalization of a many-body quantum system is one of the major areas of research of the last few decades in quantum and statistical physics \cite{Mori_2018}. A simplified definition of the thermalization phenomena is given as the complete loss of initial memory of the many-body quantum systems where the ergodicity condition holds \cite{Nandkishore2015}. Two different approaches are developed throughout the year to explain it, i.e., the first one is for the isolated quantum systems, where such systems thermalize due to their internal dynamics, whereas the second one is strictly for the open quantum systems, where the system is coupled with a comparatively larger subsystem, which is already in thermal equilibrium. Here, we mostly focus on the latter case, where it is expected that the system will reach the thermal Gibbs state at a late time and the final temperature of the system will be equal to the bath temperature. In the weak-coupling approximation and the Markovian limit, such an irreversible journey of the system is well explained by the Lindblad equation \cite{breuer2002}.  The dynamics beyond the Markovian regime and the weak-coupling approximation have been explored recently to understand the other demanding aspects of the irreversible dynamics \cite{Vega2017}. 

An exception can be observed when a many-body system fails to thermalize. In such cases, the initial memory of the initial states can be kept for an arbitrary long time, e.g., integrable systems \cite{kollar2011, Vidmar_2016, Akemann2019}. It has been shown that the quantum dynamics is constrained by several conserved quantities that help the system to skip thermalization. Such nonthermality has been studied thoroughly in condensed matter physics, atomic physics, and quantum information \cite{Abanin2019}. A different class of systems that shows striking features of nonthermality is the nearly integrable quantum systems \cite{Langen_2016, kollar2011, Saha2023, Saha_2024, Mallayya2019}. There exists a clear time-scale separation in the dynamics of such non-integrable systems. However, the extremely long-time dynamics have no difference from the thermalization. But in the intermediate time scale, the system reaches a quasi-stationary state, which is widely known as the prethermal state. The dynamics is constrained by several quasi-conserved quantities in this regime. Recently, it was obtained that an open quantum system consisting of non-interacting spins can show the notion of prethermalization in the presence of an engineered reservoir with a long bath correlation length \cite{Saha_2024}. Such an environment with a high correlation length can hold the quantum memory for a longer time, which ultimately leads to prethermalization. There also exist several states that act as a decoherence-free subspace in the prethermal regime. Such a state has several applications in making an efficient entanglement storage device, which was explored recently \cite{Saha_2022}.  Motivated by the above-mentioned ideas, we want to study the prethermal behavior of open quantum systems in a different context, i.e., the Unruh effect.

The indistinguishability between the quantum vacuum fluctuation and the thermal fluctuation leads to the discovery of the Unruh effect \cite{Unruh_1976}. Such a connection between the quantum vacuum and the thermal bath is first introduced as the thermalization theorem by Fulling and Davies, which tells that the inertial vacuum acts as a thermal bath for a uniformly
accelerated observer \cite{Fulling_1973, Davies_1975}. Moreover, it was shown that the temperature of the bath is proportional to the
acceleration of the detector. Later, Unruh put this theorem in the context
of particle creation, which, after him, is known as the renowned Unruh effect \cite{takagi1986vacuum}. It is well known that the Unruh thermalization is also connected to the Hawking radiation in the near-horizon region \cite{hawking1974black}.

From the perspective of a uniformly accelerated atom prepared in an initial pure state, it interacts with the vacuum state of a free scalar field and goes through a non-unitary time evolution. As a result, the system will reach a thermal mixed state at equilibrium. Following the prescription of Benatti \etal, an open quantum system approach of the atoms where the effective temperature of the bath is
proportional to the acceleration ($\alpha$) of the atoms \cite{Benatti_2002}. Hence, the Lindblad equation successfully captures the dynamics of the system. The steady-state solution of the Lindblad equation is the thermal state, which is a mixed state with the maximum von Neumann entropy.  The initial
memory of the atom is completely lost during this irreversible time evolution, and no conserved quantities survive at thermal equilibrium.

However, more interesting features can be observed when we increase the number of atoms in this system. For example, the emergence of entanglement between two non-interacting accelerated atoms occurs for a particular regime of inter-atomic distance and acceleration \cite{Benatti_2002}. We note that in such a regime, the thermal bath acts as a common bath which creates quantum correlations between the atoms \cite{Benatti2003}. An inter-atomic distance-dependent thermal-nonthermal signature of the Unruh effect in resonance Casimir-Polder interaction (RCPI) was reported earlier. At longer atomic separation, the local inertial approximation is not valid, hence the system shows non-thermal behavior. On the other hand, such a response is thermal at lower separation due to the validity of local inertial approximation \cite{Marino2014, Rizzuto2016, Singha2020, Saha2021, Chatterjee_2020, Tian2014}. We argue that in this regime of local inertia, the system may skip the Unruh thermalization, as the dynamics are constrained by several conserved quantities. Here, we explore such a regime, where the time evolution is different from Unruh thermalization. 

The significant similarities between the common-environment problem and entanglement generation due to vacuum fluctuation bring our major interest in investigating the existence of an Unruh-prethermal state for a uniformly accelerating two-atom system \cite{Benatti2003, Benatti_2002, Saha_2024}. We also extend our analysis to the many-atoms case to observe whether such a system shows the notion of collective behavior or not. A set of non-interacting atoms initially prepared in an excited state shows the notion of Dicke superradiance due to coupling with the common modes of the cavity \cite{dicke_coherence_1954}. One key feature of the Dicke superradiance is that the decay profile of the spontaneous emission is non-monotonic in nature, and the radiation burst is more intense and short-lived for increasing the number of atoms \cite{gross_superradiance_1982, Masson2022, Saha_2025}. One of our central results in this paper is the similarities between the intermediate scale dynamics towards the Unruh prethermal state of a many-body quantum system and the Dicke superradiance.

 In this manuscript, we confine ourselves to the Markovian regime. We note that the source of non-thermality, which is investigated here, is completely different from the non-Markovian Unruh effect \cite{Sokolov2020}.  We also use the exact experimental values estimated by Bell and Leinass in our numerical simulation to provide an estimation of the prethermal regime \cite{Bell1983}. Such an estimation will be helpful for future experiments involving the observation of Unruh prethermalization.

Throughout the paper, we follow
the given notation. The four-vector is denoted by `$x$,' the time is
represented by `$t$', and the proper time by `$\tau$.' We denote the space co-ordinate by `${\bf{x}} $,' $ ({\bf{x}} = X,Y,Z)$. We arrange the manuscript in the following order. In Sec-\ref{sec-ii}, we provide a brief description of the system where we explicitly show the vacuum correlation function and the analytical form of the Lindblad equation. We consider the two-atom problem in Sec-\ref{sec-iii}, where in terms of the two-atom observable, we calculate the steady state and predict the emergence of a prethermal state, and we also show the entanglement generation as a feature of the prethermal state. In Sec-\ref{sec-iv}, we extend our approach to the many-atom case, where we particularly focus on two aspects: the von Neumann Entropy for the Unruh prethermal state and the Dicke-type short-lived radiation burst in intermediate-scale dynamics. We also compare our results with the existing experimental works in Sec-\ref{sec-v}. Finally, we discuss the other aspects of our aspects in Sec-\ref{sec-vi} and conclude in Sec-\ref{sec-vii}.

\section{Dynamics of two accelerated atoms through a massless scalar field}
\label{sec-ii}
Here, we consider a system 
of two atoms (two-level systems) uniformly accelerated in the
Minkowski spacetime interacts with a free massless scalar field. The atoms
are following the hyperbolic trajectory with respect to the inertial observer.
The positions of the atoms in terms of proper time are written by,
\begin{eqnarray}
&&t_a =\frac{1}{\alpha} \sinh \alpha  \tau,\,  X_a=\frac{1}{\alpha} \cosh \alpha \tau, \, Y_a= 0,\,\,Z_a =Z_{a}  \nonumber\\
&& t_b =\frac{1}{\alpha} \sinh \alpha  \tau,\,  X_b=\frac{1}{\alpha} \cosh \alpha \tau, \, Y_b= 0, \,Z_a =Z_{b}.\nonumber\\
\label{eq:1}
\end{eqnarray}
We define $L$ to be the proper distance between the two atoms located at the positions 
$(t_{a}(\tau),\, \bf{x}_{a}(\tau))$ and $(t_{b}(\tau),\bf{x}_{b}(\tau))$.

The Hamiltonian of the system + scalar field is expressed as,
\begin{eqnarray}
\Hq=\hs^\circ+\hl^\circ+\hsl~.
\label{full_hamiltonian}
\end{eqnarray}
Throughout the paper, we use the natural units, $\hbar=c=1$. Here, we assume the
two atoms have the same Zeeman levels, and the free Hamiltonian of the system is
$\hs^\circ=\hs^{\circ 1}+\hs^{\circ 2}=\frac{1}{2}\omega_\circ \sigma_3^{(1)}
+\frac{1}{2}\omega_\circ \sigma_3^{(2)}$ (The superscript in the Pauli matrices represents the atom
number). $\sigma_i$ is the Pauli matrix, and $\omega_\circ$ is the frequency of
Zeeman levels. The free Hamiltonian of the scalar field $\hl^\circ$ can be
expressed as, $\hl^\circ=\int \frac{d^3 k}{(2\pi)^3}~\omega_k a^{\dagger}(k) a(k)~.$
Here, $\omega_k $ is the frequency of the free scalar field, and $a^{\dagger},
a$ are the creation and annihilation operators of the quantized field,
respectively. The coupling Hamiltonian between the atom and the scalar field is given by \cite{Benatti_2002},
\begin{eqnarray}
\hsl = \lambda \sum\limits_{\mu=0}^{3} \big[\sigma_{\mu}^{(1)} \otimes \phi_{\mu}(x_1)   
+ \sigma_{\mu}^{(2)} \otimes \phi_{\mu} (x_2)\big]~,
\label{coupling_hamiltonian}
\end{eqnarray} 
where $\lambda$ is the coupling constant, $\phi$ represents the scalar field
and $x_{1},\,x_2$ are the individual trajectories of two atoms. 
Here, $\phi_{\mu} (x)=\sum^N_{a=1}\big[\chi^a_{\mu} \phi^-(x)\,
+\,(\chi^a_{\mu})^{\dagger} \phi^+(x)  \big]~$.
$\phi^{\pm}(x)$ is the positive and negative field operator of the
free scalar field and $\chi^a_{\mu}$ are the corresponding complex
coefficients. It is assumed that the spins and the scalar field are initially
uncorrelated, so the total initial density matrix can be written as $\rho(0)=
\rho_s(0) \otimes \vert 0\rangle \langle 0 \vert$. $\vert 0 \rangle $ is the
vacuum state of the scalar field and $\rho_s(0) $ is the initial density matrix
of the system, such an assumption is consistent with the Born-Markov approximation \cite{breuer2002}. We are also working in the weak system-scalar field-coupling limit. Using the von Neumann Liouville
equation in the interaction frame, expand it up to the second-order terms of $\hsl$  and finally take the trace over the field variable, we get the dynamical equation of the system, which is written below,
\begin{eqnarray}
\frac{d \rho_s(\tau)}{d\tau}= -i\Big[\mathcal{H}_{lamb},\,\rho_s(\tau)\Big]+\mathcal{L}\big(\rho_s(\tau)\big)~. 
\label{eq:4}
\end{eqnarray}
The above Eq. (\ref{eq:4}) is known as the Lindblad quantum master equation (QME) \cite{breuer2002}. Here,
$\mathcal{L}\big(\rho_s(\tau)\big)$ is the dissipator of the master-equation
and  $\mathcal{H}_{lamb}$ is known as the Lamb-shift Hamiltonian that
leads to the renormalization of the Zeeman Hamiltonian. We note that the Lamb shift
gives rise to the CPI between the spins \cite{Saha2021, Chatterjee_2020}. The analytical forms
are given by, 
\begin{eqnarray}
\mathcal{L}\big(\rho_s\big) &=& \sum \limits_{a,b=1}^{2} \sum \limits_{j,k=1}^3 \gamma^{ab}_{jk}\Big(\sigma _b ^k\rho_s \sigma _a ^j  
-\frac{1}{2}\{ \sigma _a ^j \sigma _b ^k ,\rho_s \} \Big)~,\\
\label{eq:5}
\mathcal{H}_{lamb}&=& -\frac{i}{2} \sum \limits_{a,b=1}^{2} \sum \limits_{j,k=1}^3 \mathcal{S}_{jk}^{ab} \sigma _a ^j \sigma _b ^k~.
\label{eq:6}
\end{eqnarray}
The field correlation time-scale is taken to be smaller than the relaxation
time-scale of the system. The dynamics are completely positive, and trace preservation
holds (CPTP condition). Here, we assume that $\chi^a_{\mu} $ satisfies
$\sum\limits^N_{a=1} \, \chi^a_{\mu} \big( \chi^a_{\nu}\big)^{\dagger}\,=\,
\delta _{\mu \nu}.$ Hence, the field correlation functions are diagonal, i.e.
$G_{ij}^{ab}(x-y)= \delta_{ij}G^{ab}(x-y)$. The expressions of the
field-correlation function are obtained by,
\begin{eqnarray}
 G^{a b}(\Delta \tau)&=&\langle 0 \vert \hat{\Phi} (\tau,{\bf{x}}_a) \hat{\Phi}(\tau^{\prime},{\bf{x}}_b)\vert 0\rangle~.
\label{correlation}
\end{eqnarray}
Here $\Delta \tau = (\tau-\tau')$. The correlation function is  also assumed to be
stationary, which is an essential condition for the Markovian time-evolution. For a massless scalar field, in the four-dimensional Minkowski spacetime, the expression of this function is written as \cite{takagi1986vacuum}, 
\begin{eqnarray}
G(x,x')=-\frac{1}{4\pi^2 R^2}~,
\end{eqnarray}
where $R = \sqrt{(t-t^{\prime}-i \epsilon)^2-|{\bf{x}}-{\bf{x}}^{\prime}|^2}~$.
It is called the positive-frequency Wightman function. Here \,$i \epsilon$\,
is a small constant in the complex plane and acts as a regulator to avoid the
divergence and make sure about the function's analytic property over the chosen
complex plane.  The Fourier transform of the correlation function along the trajectory of the atoms is
given below,
\begin{eqnarray}\label{resp}
\mathcal{G}^{ab} (\omega_\circ) &=& \frac{1}{2\pi}\frac{\omega_\circ}{1-e^{- 2\pi\omega_\circ /\alpha}}f_{ab}(\alpha)~. 
\end{eqnarray} 
Here, $ f_{ab}(\alpha)=\frac{\sin\Big(2\,\frac{\omega_\circ}{\alpha}\sinh^{-1}(\alpha \, L/2)\Big)}{L \,\omega_\circ \sqrt{1+L^2\, \alpha^2/4}}$ for $a\neq b$, and $f_{ab}(\alpha)=1$ for $a = b$.  
Using the above Eq. (\ref{resp}), the explicit form of the Kossakowski matrix, which is an essential component
of the Lindbladian, is given by \cite{Benatti_2002},
\begin{eqnarray}
\gamma^{ab}_{jk}= A^{ab}\delta_{jk}-i B^{ab}\epsilon_{jkl} \delta_{3l}-A^{ab}\delta_{3k} \delta_{3l}~.
\end{eqnarray}
 $A^{ab}\,  (B^{ab})$ are the even (odd) combination of $\mathcal{{G}}^{ab}$ for the positive and negative Zeeman frequency.

For a massless scalar field, the explicit forms of $A^{ab},  B^{ab}$ are written as, 
\begin{eqnarray}
B^{ab}=\frac{\lambda^2 \,\omega_\circ}{8\pi} f_{ab}(\alpha),\, A^{ab}= B^{ab}\times \coth (\pi\omega_\circ/\alpha). 
\end{eqnarray}
Here $a,b=\{1,2\}$. We note that $f_{ab}$ has two extreme limits based on the value of $\alpha L$. For $L \ll 1/\alpha$, $f_{ab} \approx 1$, and for $L \gg 1/\alpha$, $f_{ab} \approx 0$. In terms of (up)downward transition rate $(\gamma_+)\gamma_-$, the Lindblad equation is written as,
\begin{eqnarray}
\mathcal{L}(\rho_s ) &=& \sum \limits_{a,b=1}^{2} \gamma_+\Big(\sigma_+ ^a\rho_s \sigma_- ^b  
-\frac{1}{2}\{ \sigma_- ^b \sigma_+ ^a ,\rho_s \} \Big)~\nonumber\\
&& + \gamma_-\Big(\sigma_- ^a\rho_s \sigma_+ ^b  
-\frac{1}{2}\{ \sigma_+ ^b \sigma_- ^a ,\rho_s \} \Big) 
\end{eqnarray}
Here, $\gamma_+ = \frac{\lambda^2 \,\omega_\circ}{8\pi}(1 + \tanh (\pi\omega_\circ/\alpha))/2$, and $\gamma_- = \frac{\lambda^2 \,\omega_\circ}{8\pi}( 1 - \tanh (\pi\omega_\circ/\alpha ))/2$.

\section{Dynamical equation of the system}
\label{sec-iii}
The steady-state solution of the QME gives the notion of the equilibrium state of the system. In the Liouville space, the QME can be written as,
$\frac{d}{d\tau}\hat{\rho_s}= \hat{\mathcal{L}} \hat{\rho_s}$. Here, $\hat{\rho_s}$ is a
column vector of dimension $N^2\times 1$ and $\hat{\mathcal{L}}$ is the
Lindbladian superoperator of dimension $N^2 \times N^2$ ($N$ is the dimension of the Hilbert space). The eigenvector
corresponding to the zero eigenvalue is the steady-state solution of the QME.
In our case, the steady-state solution depends on the behavior of $f_{ab}$. For a single
spin-$1/2$ system and $2 \times 2$ matrix, the observables are
$(\sigma_x,\sigma_y,\sigma_z)$ (one constraint, i.e., the trace preservation) and the corresponding equations are known as the Bloch equations. Similarly, for two spin systems, there exist fifteen observables. As the Zeeman levels of the spins are the same, the number of observables is further reduced to nine, which can be written as \cite{Saha2023},
\begin{eqnarray}\label{operator}
M_i &=& \frac{1}{2}\tr_s \Big([\sigma_i^{(1)} \otimes \mathbb{I}+\mathbb{I} \otimes \sigma_i^{(2)}]\rho_s \Big)~,\nn\\
M_{ii} &=& \frac{1}{4}\tr_s \Big([\sigma_i^{(1)}\otimes \sigma_i^{(2)}] \rho_s \Big)~,\nn\\
M_{ij} &=& \frac{1}{4}\tr_s \Big([\sigma_i^{(1)} \otimes \sigma_j^{(2)}+\sigma_j^{(1)} \otimes \sigma_i^{(2)}] \rho_s \Big)~.
\end{eqnarray}
 The Bloch-type equations for our case are written as,
\begin{equation}\label{3eq}
\left[\begin{array}{c} \dot{M}_z \\ \dot{M}_{zz}\\ \dot{M}_{c} \end{array} \right] = \begin{bmatrix}
-A^{11}  & 0 & 2B^{12} \\  B^{11}/2& -2A^{11}  & A^{12} \\ 
-B^{12}/2 & 2A^{12}  & -A^{11}   \end{bmatrix}  \left[\begin{array}{c} M_z  \\
M_{zz}\\ M_c \end{array}\right] + \left[\begin{array}{c} B^{11} \\  0\\ 0 \end{array}\right]~.  
\end{equation} 
Here $M_c = M_{xx}+M_{yy}$ and $f_{ab}$ varies from $0 \leq f \leq 1$. We only present the equation of observables related to the population of the density matrix. In the next subsection, we will show the steady state solution for different values of $f_{ab}$.

\subsection{Steady state solution for $0\leqslant f_{ab}<1$}
\begin{figure}[!htbp]
    \includegraphics[width=\linewidth]{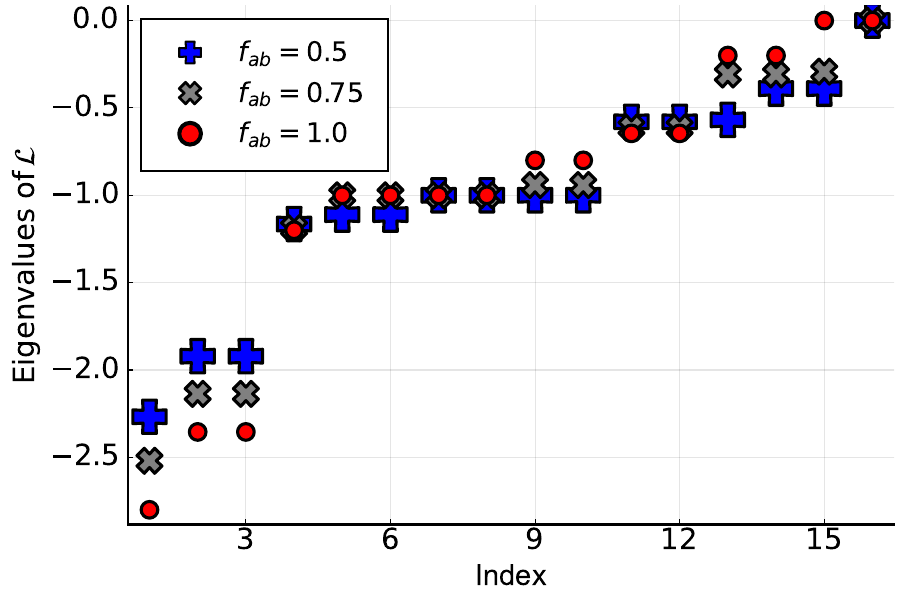}
    \caption{The plot of the eigen-spectrum of the Liouvillian for different values of $f_{ab}$. The effect of $\mathcal{H}_{lamb} $  is neglected, hence all the sixteen eigenvalues of $\mathcal{L}$ for the two-spin case are real. For $f_{ab} = 1$, we have two zero eigenvalues, whereas for $f_{ab}<1$, we have a single zero eigenvalue. With decreasing $f_{ab}$, the fifteenth eigenvalue also decreases. Hence, at $f_{ab} \to 1$, it is very close to 1 and hence shows the notion of prethermalization. Here, we choose $\gamma_+ = 0.8$, and $\gamma_- = 0.2$.}
        \label{fig:spectrum}
\end{figure}

We note that, in this regime, the Lindbladian has a single zero eigenvalue, which is shown in Fig. \ref{fig:spectrum}. We note that the steady-state solution is independent of the initial value dependency. Such a solution is given by,
\begin{eqnarray}
M_z^{eq}&=& M_\circ~,\nn\\
M_{zz}^{eq} &=& \frac{M_\circ^2}{4},\nn\\
M_c^{eq}&=&0~.
\end{eqnarray}  
Using the above solution, the steady-state density matrix is given by,
\begin{eqnarray}
    \rh^{eq} &=& \frac{e^{- \frac{\pi}{ \alpha} \hs^\circ}} {\mathcal{Z}} \nonumber\\
    &=& \mathbb{I} /4 +\frac{M_z^{eq}}{2} ( I_z^1 + I_z^2) + 4M_{zz}^{eq} I_z^1 I_z^2
\end{eqnarray}
Here, $\mathcal{Z}$ is the partition function, and $I_i = \sigma_i/2$. In this regime, the expression for $\rh^{eq}$ indicates that the thermalization theorem holds as the final state is a thermal Gibbs state \cite{takagi1986vacuum}.
\subsection{Steady state solution for $f_{ab} = 1$}
Before discussing the steady state solution for this case, we give a short review of the symmetries and conserved quantities in open quantum systems \cite{Lieu2020, Albert2014}. Symmetry can be classified into two categories. If an operator $\mathcal{O}$ commutes with both the Hamiltonian and the jump operators of the Liouvillian, then it is defined as the strong symmetry operator. On the other hand, if the symmetry super-operator only commutes with the total Liouvillian, not the individual components (i.e, Hamiltonian and jump operators), then it is called the weak symmetry. Hence, the strong symmetry conditions already satisfy the weak symmetry condition, but the reverse condition is not always true.

In our case, we find that at $f_{ab}=1$, the Liouvillian has two zero eigenvalues (shown in Fig. \ref{fig:spectrum}), which corresponds to the existence of another symmetry or conserved quantities apart from the trace preservation. 
Here, $A^{11}=A^{12}$, $B^{11}=B^{12}$. The Lindbladian super-operator
$\hat{\mathcal{L}}$ is invariant under the following unitary transformation,
$\hat{U}(\kappa)\hat{\mathcal{L}}\hat{U}^{\dagger}(\kappa)=\hat{\mathcal{L}}$ where
$\kappa$ is real. We define, $\hat{U}(\kappa)=\exp{[-i \hat{D} \kappa]}$ and $\hat{D}= D \otimes \mathbb{I}-
\mathbb{I}\otimes  D^T$. Here, $D= \vec{\sigma}_i.\vec{\sigma}_j$ . 
$\hat{D}$ is the symmetry super-operator. The analytical form of conserved quantity is written as,
\begin{eqnarray}
\frac{d}{d\tau}(M_{xx}+M_{yy}+M_{zz})=0~.
\end{eqnarray}  
The steady state solution in terms of observables is given by,
\begin{eqnarray}\label{newval}
M_z^{eq}&=& \frac{M_\circ(3+4(M_2+ M_3))}{3+M_\circ^2}~,\nn\\
M_{c}^{eq}&=& -\frac{M_\circ^2-4(M_2+M_3)}{2(3+M_\circ^2)}~,\nn\\ M_{zz}^{eq} &=& M_2 + M_3 -  M_{c}^{eq}.
\end{eqnarray}
Here, $M_{c}(0)=M_2$, $M_{zz}( 0)=M_3$ and $M_\circ=\tanh (\pi \omega_{\circ}/ \alpha)$. We note that the steady state solution has the initial value dependency, which is a sharp contrast from the thermal Gibbs state. The steady state density matrix is given by,
\begin{eqnarray}
    \rho_s^{eq} &=& e^{-\frac{\pi }{\alpha}\hs^\circ - \frac{l_1}{4} \vec{\sigma_1}.\vec{\sigma_2}}/\mathcal{Z}_g \nonumber\\
    &=& \mathbb{I}/4  + \frac{M_z^{eq}}{2} ( I_z^1 + I_z^2)+ 4M_{zz}^{eq} I_z^1 I_z^2 \nonumber\\
    &+& 2M_c^{eq}(I_x^1 I_x^2  + I_y^1 I_y^2 )\\
    \text{here}\quad l_1 &=& \ln \bigg( \frac{1 - 4(M_2 +  M_3)}{3 + 4(M_2 +  M_3) }   (1 + 2 \cosh \frac{\pi \omega_\circ}{\alpha}) \bigg) \nonumber
    \end{eqnarray}
    $\mathcal{Z}_g$ is the partition function for the generalized Gibbs state.
   \begin{figure*}[!htbp]
\subfigure[]{\includegraphics[width=0.3\textwidth]{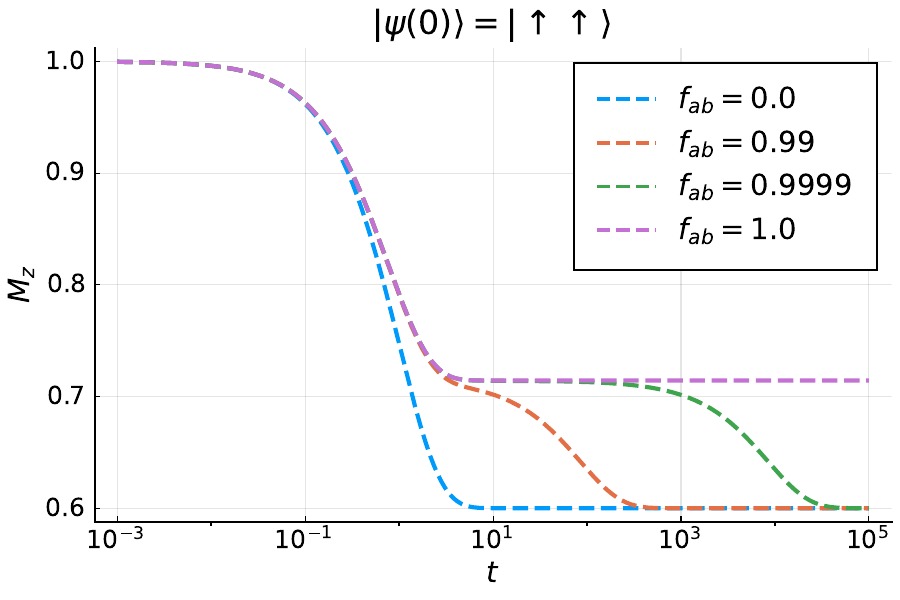}}
\subfigure[]{\includegraphics[width=0.3\textwidth]{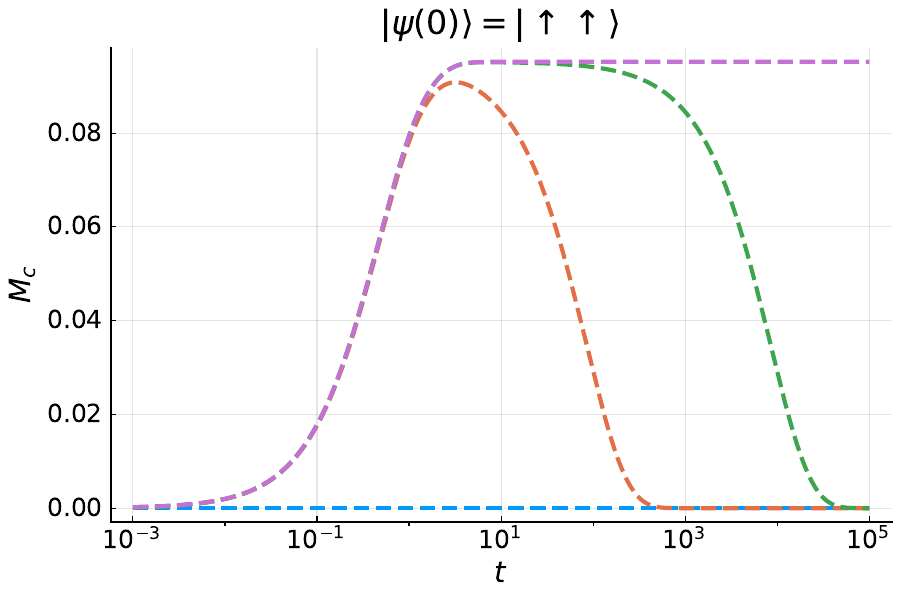}}
\subfigure[]{\includegraphics[width=0.3\textwidth]{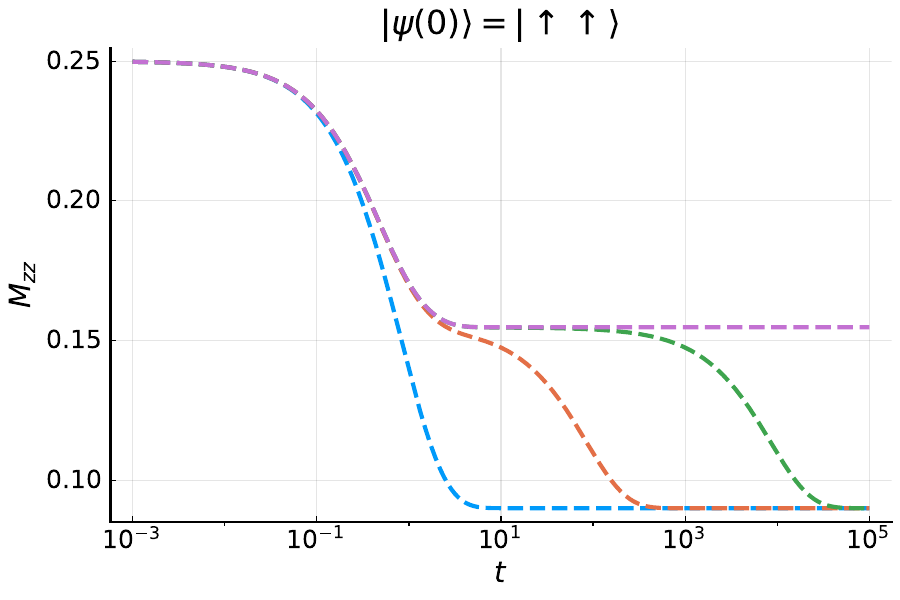}}
\subfigure[]{\includegraphics[width=0.3\textwidth]{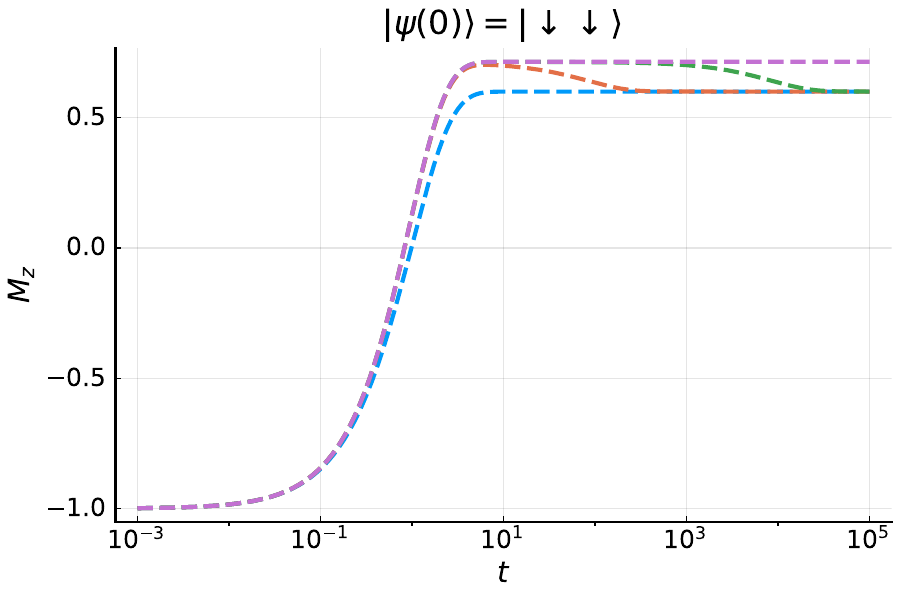}}
\subfigure[]{\includegraphics[width=0.3\textwidth]{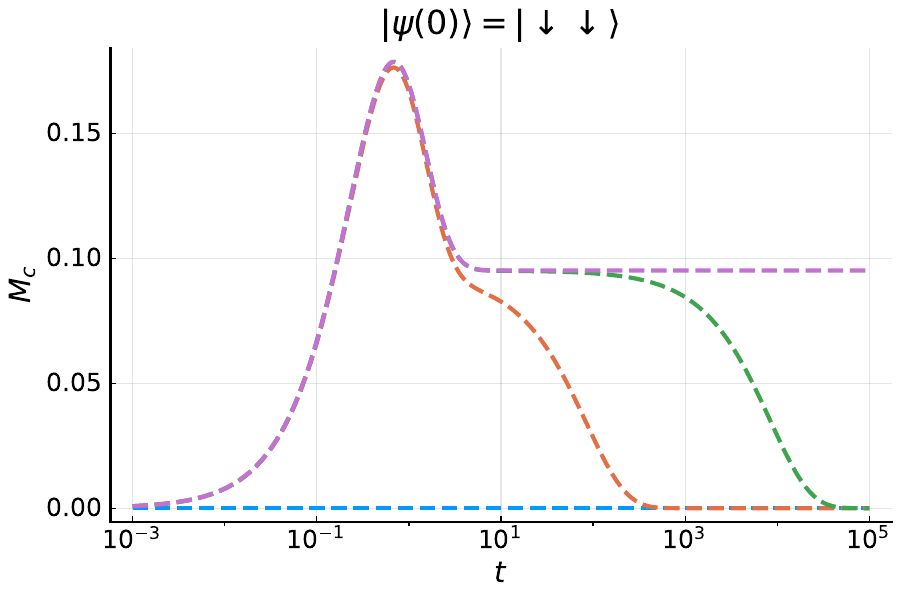}}
\subfigure[]{\includegraphics[width=0.3\textwidth]{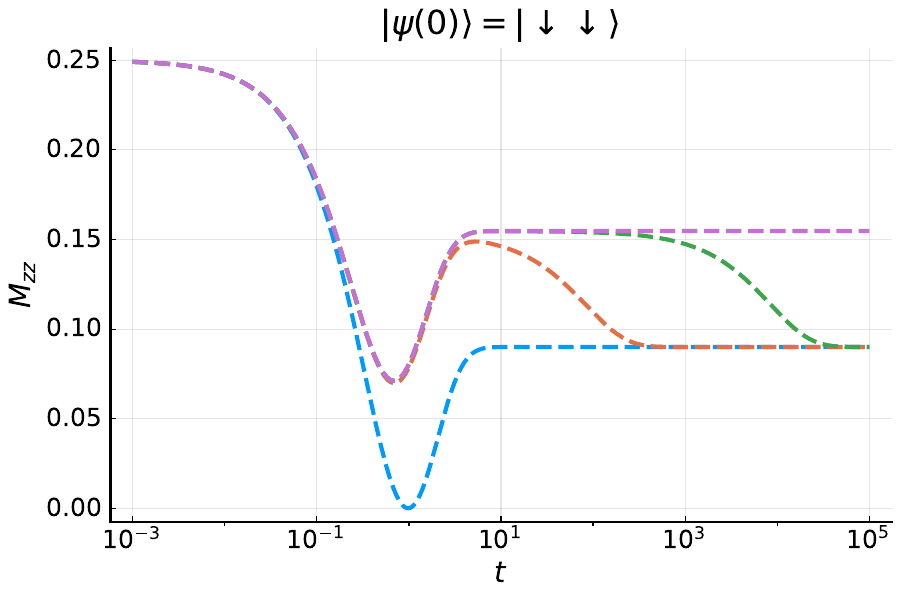}}
\subfigure[]{\includegraphics[width=0.3\textwidth]{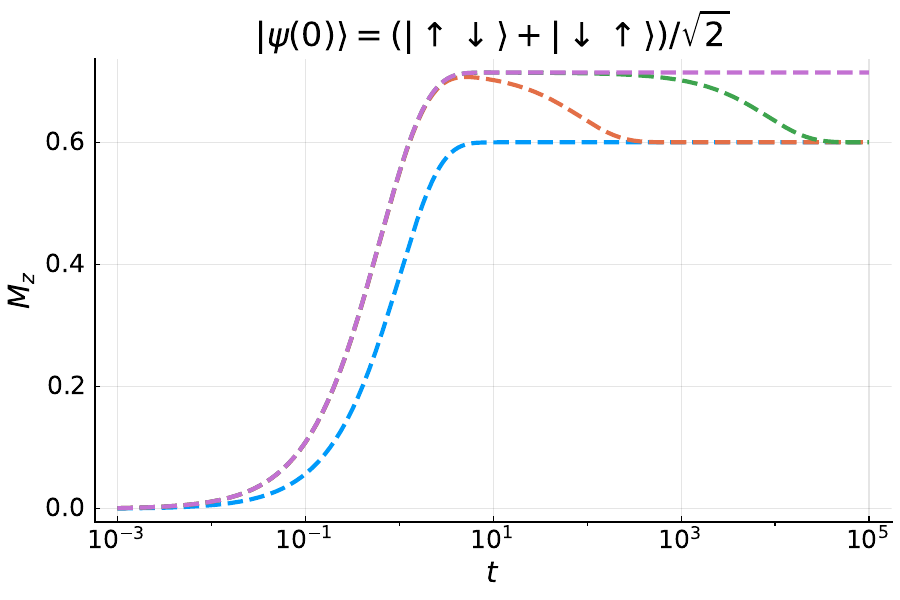}}
\subfigure[]{\includegraphics[width=0.3\textwidth]{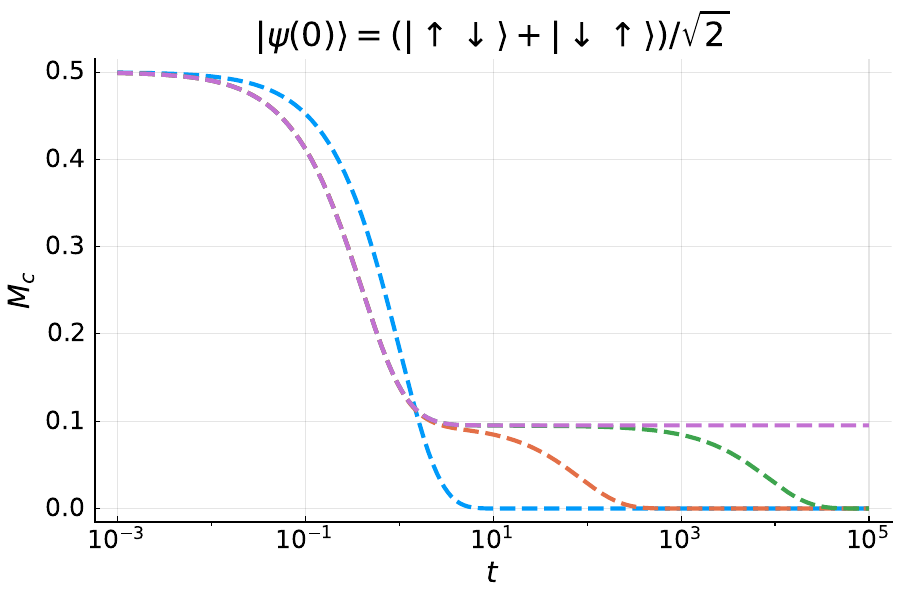}}
\subfigure[]{\includegraphics[width=0.3\textwidth]{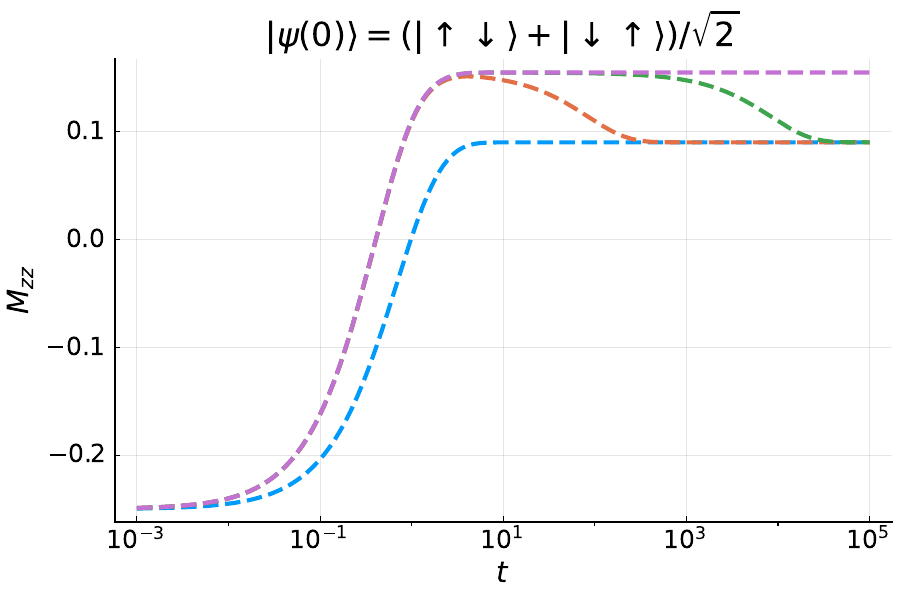}}
\subfigure[]{\includegraphics[width=0.3\textwidth]{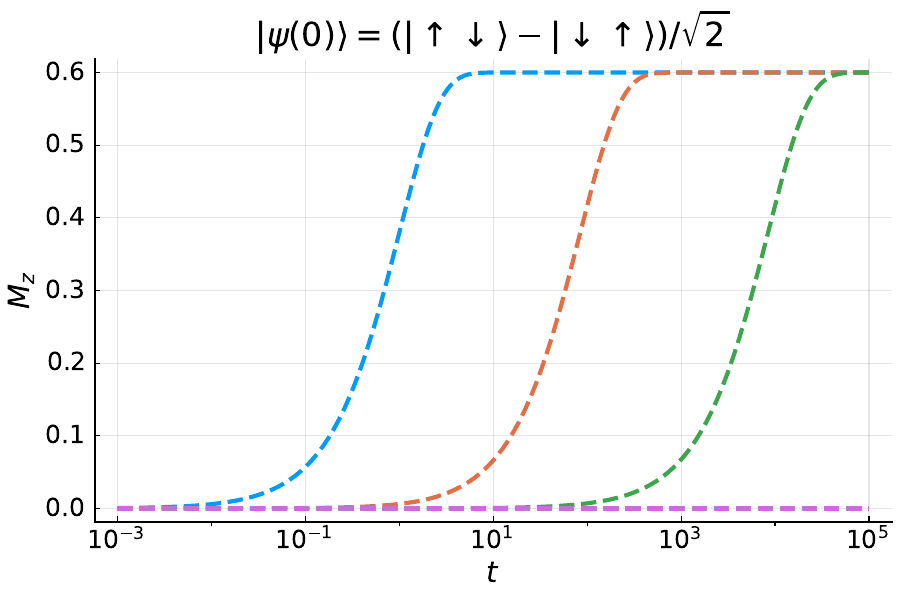}}
\subfigure[]{\includegraphics[width=0.3\textwidth]{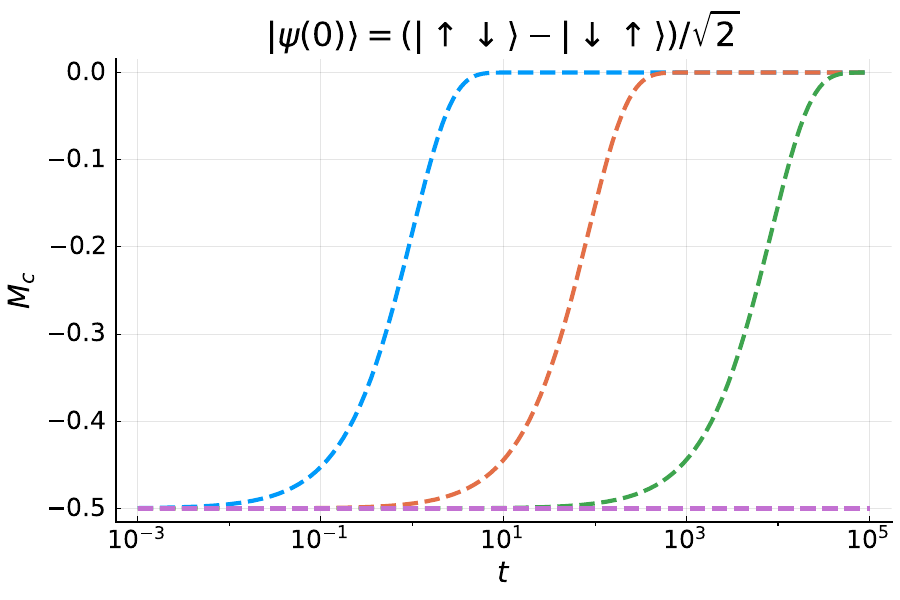}}
\subfigure[]{\includegraphics[width=0.3\textwidth]{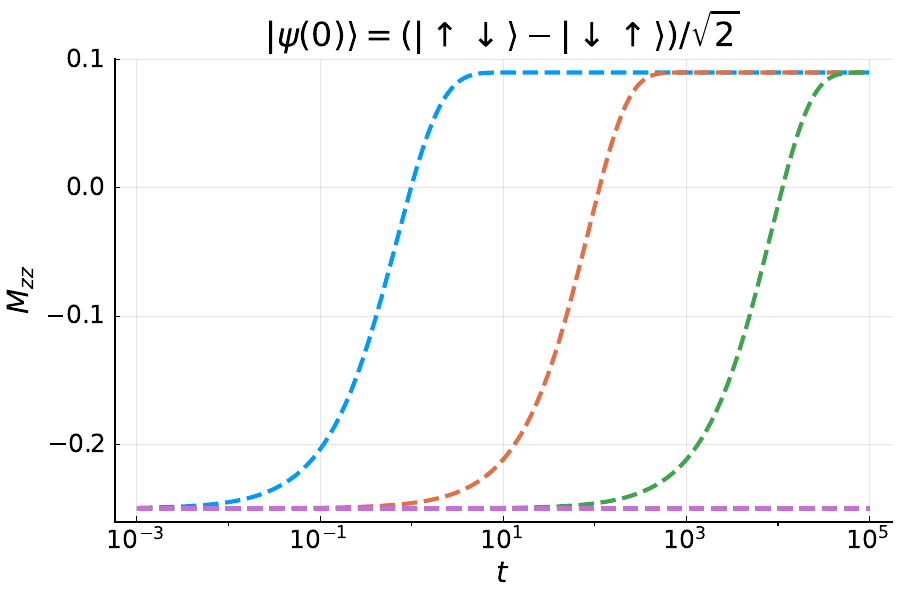}}
\caption{The plot of the time-evolution of the relevant observables $\{M_z, M_{zz}, M_c\}$ for different choices of $f_{ab}$ and initial condition. Here $N=2$. The initial states are chosen from the dipolar basis as it is the eigen-basis of the symmetry operator $\vec{\sigma_1}. \vec{\sigma_2}$. We note that the steady state value of the observables is different for $f_{ab} =1$ and $f_{ab}<1$. The steady state for $f_{ab} <1$ is the Gibbs state, and for $f_{ab} =1$ it is the generalized Gibbs state. For $f_{ab} \to 1$, the system spends a longer time at a quasi-steady state, which is the same as the steady state for $f_{ab} =1$, which shows the notion of prethermalization. The singlet state $\vert \psi \rangle = (\vert \uparrow \downarrow\rangle - \vert  \downarrow \uparrow\rangle )/ \sqrt{2}$ never evolves for $f_{ab} =1$, hence acts as a dark state of decoherence-free subspace. Here, we choose $\gamma_+ = 0.8$, and $\gamma_- = 0.2$. As our focus is to explore the Gibbs state, a long-lived prethermal state, and generalized Gibbs state, we choose the following four values of $f_{ab} = \{0.0, 0.99, 0.9999, 1.0\}$. }
    \label{fig:enter-label}
\end{figure*}

\subsection{Emergence of prethermalization at $f_{ab} \to 1$}
$f_{ab}=1$ is an asymptotic limit, as such a value is only possible when $L = 0$, or $1/\alpha  = 0$. Such values are not possible experimentally. Here, our main focus is to study the regime of high acceleration and closely spaced atoms. In this regime, for the two-atom case, the Liouvillian has a single zero eigenvalue, but the nearest eigenvalue is very close to zero. 

Following our recent work, the prethermalization can be identified by looking at the structure of the eigen-spectrum of Liouvillian, i.e., the eigen-spectrum consists of a single zero eigenvalue, and the real part of a few eigenvalues is very close to zero in comparison to the remaining sets of eigenvalues \cite{Saha_2024}.  We note that in this case, the eigen-spectrum follows a similar structure. Hence, we conclude that the system reaches a prethermal quasi-steady state before it thermalizes. We call such phenomena as the ``Unruh-prethermalization". As such, in the intermediate time-scale, the system follows the dynamics for $f_{ab}=1$, so the quasi-steady state is a generalized Gibbs state. However, the system follows the dynamics for $0\leqslant f<1$ at a later time. Hence, it reaches the thermal Gibbs state in the long-time limit. 

We are interested in the dynamics of the system in the dipolar basis, as it is the eigen-basis of the symmetry operator. The corresponding states are $\vert \uparrow \uparrow \rangle$, $\vert \downarrow \downarrow \rangle$,  $(\vert \uparrow \downarrow \rangle +\vert \downarrow \uparrow \rangle)/\sqrt{2}$, and $(\vert \uparrow \downarrow \rangle -\vert \downarrow \uparrow \rangle)/\sqrt{2}$. Among the four states, the first two states are separable and the next two are maximally entangled states. We show the plots for the relevant observables, $\{M_z, \, M_{zz},\, M_c\}$ using the above four states as an initial configuration. We only consider four values of $f_{ab}$ = $\{0.0,\, 0.99,\, 0.9999,\, 1.0\}$. Apart from the singlet state (i.e., $\vert \psi(0) = (\vert \uparrow \downarrow \rangle -\vert \downarrow \uparrow \rangle)/\sqrt{2} $), Fig. \ref{fig:enter-label}, depicts the emergence of prethermalization for $f_{ab} \to 1$, as the quasi-steady state is similar to $f_{ab} =1$, whereas, the final steady state is same as $f_{ab} =0$. We note that $1/(1-f_{ab})$ is a good identifier of the lifetime of the prethermal state. For the singlet state, there is no evolution for $f_{ab} =1$, hence it acts as a dark state. A detailed discussion is provided in the next section.

To exhibit the presence of a prethermal state, we also plot the purity of the quantum state as a function of time in Fig. \ref{fig-cp-1}. The purity is defined as $\tr(\rho_s^2)$. For a pure quantum state $\tr(\rho_s^2)=1$, whereas, for mixed state $\tr(\rho_s^2)<1$. The purity is a minimum for the thermal state. We show that for $f_{ab} \to 1$, as the system reaches the prethermal state, $\tr(\rho_s^2)$ becomes constant at that time-scale and further it decays to the minimum value.

\subsection{Entanglement  as a measure of the lifetime of prethermal state}
In case of multiple zero eigenvalues, the pure eigenvector of the corresponding eigenvalue acts as a dark state or decoherence-free subspace \cite{Buca_2012}. Hence, if we choose this state as an initial state, it never evolves. Such a state is also represented as an eigenvector of the symmetry operator \cite{Saha_2024}. In this case for $f_{ab}=1$, the dark state is $\vert \psi
\rangle=\frac{1}{\sqrt{2}}(\vert 0  1 \rangle-\vert 1 0 \rangle)$, which is also an eigen-state of $D$. It is a maximally entangled anti-symmetric Bell state. If we choose the value of $M_2=-1/4$ and $M_3=-1/2$, (which corresponds to the singlet state), the final steady state solution is the same as the initial state, ($M_z^{eq}=0$, $M_{zz}^{eq}= M_2$, $M_{c}^{eq}= M_3$). The purity of such an initial state is conserved, $[\tr_s(\rho_s ^2)=1]$. 
Presence of such entangled dark states indicates that entanglement can be generated between the non-interacting atoms at the $f_{ab} \to 1$ limit.

\subsubsection{Measure of concurrence, $C(\rh)$}
Concurrence is a good measure for the growth of entanglement for a bipartite system. Following the previous works by Wooters \etal \cite{Wootters1998}, we define the concurrence of the two-atom system $C(\rho_s)$ is given by,
\begin{eqnarray}
  C(\rho_s) &=& \max \{0, \lambda_1 - \lambda_2- \lambda_3-\lambda_4\}  
\end{eqnarray}
Here, $\lambda_i$ are the eigenvalues (i.e., arranged in decreasing order) of the non-Hermitian matrix, $ \rho_s \rho_s^\prime$. We define $\rho_s^\prime = \sigma_y \otimes \sigma_y \rho_s^\star \sigma_y \otimes \sigma_y $, where $\rho_s^\star$ is the complex conjugate of $\rho_s$ provided it is written in the Zeeman basis. 
Here, $\lambda_i$ are the eigenvalues (i.e., arranged in decreasing order) of the non-Hermitian matrix, $ \rho_s \rho_s^\prime$. If $ C(\rho_s)>0$, then the system is entangled, whereas for $ C(\rho_s)=0$, the system is in a separable state. In terms of observables, $C(\rho_s)$ is written as,
\begin{eqnarray}
  C(\rho_s) = \max \{0,2\vert M_c \vert  - \frac{1}{2} \sqrt{(1 + 4 M_{zz})^2 - 4 M_z^2}\}
\end{eqnarray}
\begin{figure*}[!htbp]
\subfigure[]{\includegraphics[width=0.45\textwidth]{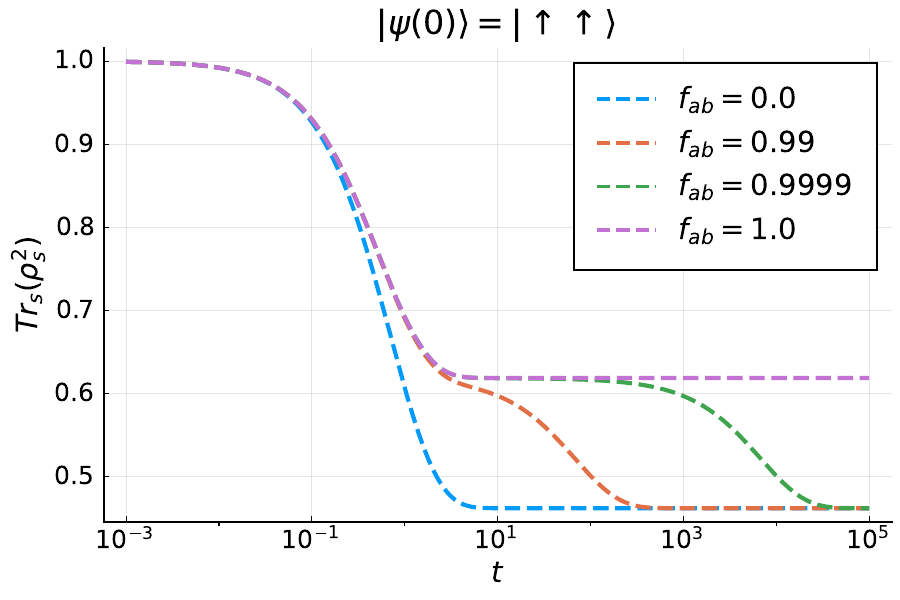}\label{fig-cp-1}}
\subfigure[]{\includegraphics[width=0.45\textwidth]{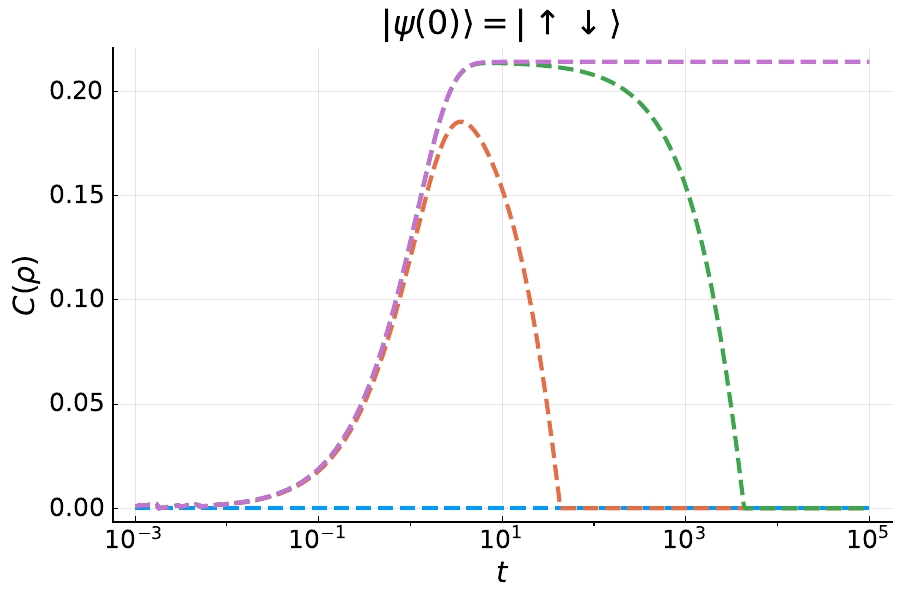}\label{fig-cp-2}}
\caption{The plot of the time-evolution of the purity operator, $\tr(\rho_s^2)$ and concurrence $C(\rho_s)$ for different choices of $f_{ab}$. Here $N=2$. For $\tr(\rho_s^2)$, we choose the initial condition as $\vert \uparrow \uparrow \rangle$. When $f_{ab} <1 $, the steady state purity is minimum, whereas, due to the presence of the symmetry, purity is increased for $f_{ab} =1 $. The system shows the notion of prethermalization for $f_{ab} \to 1$. The creation and evolution of entanglement is shown for a separable state $\vert \uparrow \downarrow \rangle$. The maximum value of $C(\rho_s)$ occurs for $f_{ab}=1$. The entanglement decays as the prethermal state evolves to the thermal state. Hence, the value of $C(\rho_s)$ is a good identifier of the life-time of the prethermal state. Here, we choose $\gamma_+ = 0.8$, and $\gamma_- = 0.2$.  } 
\label{fig-cp}
\end{figure*}
 For the singlet state, we note that the values of the observables follow the condition of $C(\rho_s) >0$. Hence, the entanglement is preserved under the evolution. We also show the plot for the creation of entanglement as a distinct feature of Unruh prethermalization. We choose a separable state ($\vert \psi(0) \rangle = \vert \uparrow \downarrow \rangle$) for the initial configuration in Fig \ref{fig-cp-2}. Our simulation shows that for the prethermal state,  the entanglement is created in the intermediate time-scale and preserved for a long time until it decays to the Gibbs state, and the lifetime increases as $f_{ab}$ gets closer to 1. After a very long time, it further decays. Hence, we claim that entanglement is a good measure for the lifetime of the prethermal state, as the decay of the prethermal state in Fig. \ref{fig-cp-1} is quite similar to the decay of entanglement in Fig. \ref{fig-cp-2}. We note that such analysis is in line with the formation of an entangled state in the presence of a common environment \cite{Landulfo2009, Benatti_2002, Wu_2022}. The value of $C(\rho_s)$ in the prethermal state is less than one, which ensures that a mixed state emerges due to interaction with the vacuum. 

\section{Extension to many-atoms case}
\label{sec-iv}
We further extend our analysis to many-atom cases by considering here a bunch of $N$ non-interacting atoms accelerating together. We only focus on the extreme limits of $f_{ab} = 0, 1$. We assume the coupling constant for atom-field interaction is the same for all atoms, which helps us to simulate the dynamics in the collective basis for $f_{ab} =1$. 

\subsection{Measure of von Neumann entropy}
Along with purity, the von Neumann entropy is also a good measure for analyzing the mixing of a system due to irreversible dynamics. The expression for the von Neumann entropy is given as, $S = -\tr(\rho_s \ln \rho_s)$. In the steady state, the von Neumann entropy becomes maximum. In terms of the partition function, such a quantity is also written as $S = -\beta^2 \frac{ \partial}{\partial \beta} (\ln \mathcal{Z}/ \beta)$. Here $\beta = \pi/\alpha$.

\begin{figure*}[!htbp]
\subfigure[]{\includegraphics[width=0.45\textwidth]{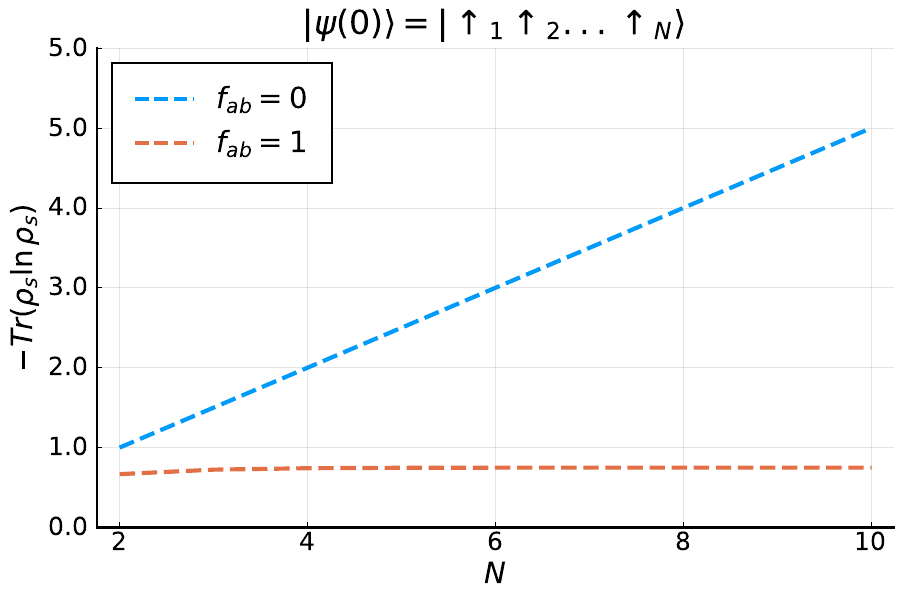}\label{fig:entropy}}
\subfigure[]{\includegraphics[width=0.45\textwidth]{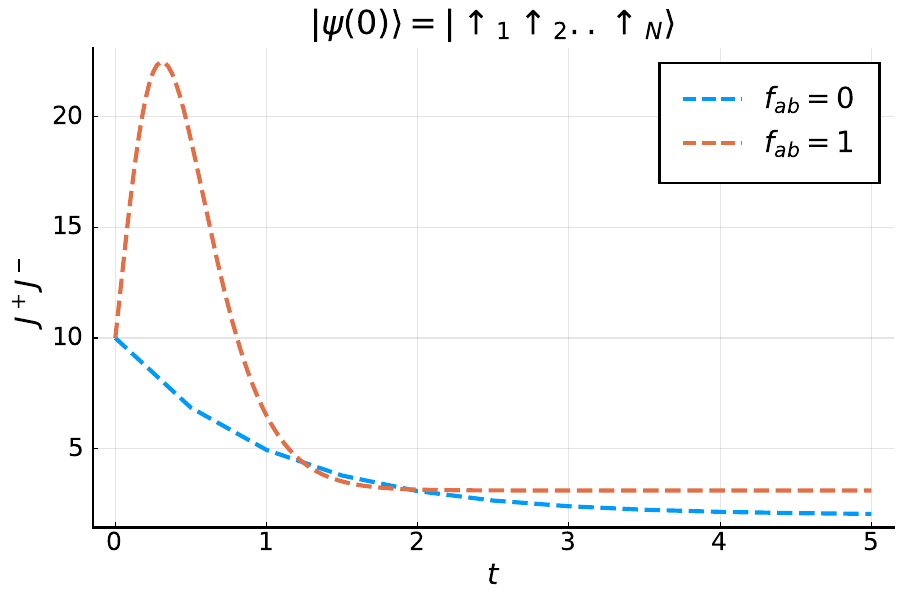}\label{fig:dicke_entropy}}
\subfigure[]{\includegraphics[width=0.45\textwidth]{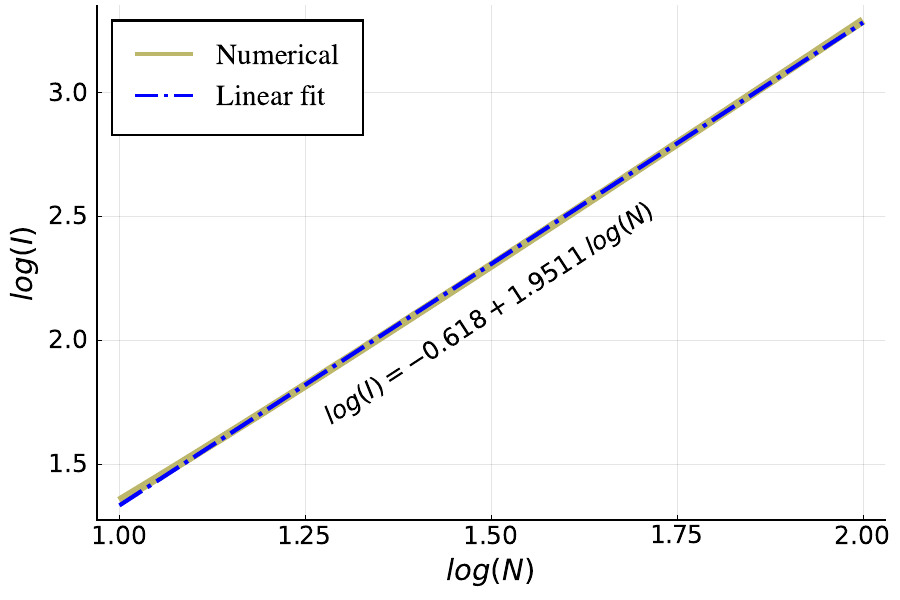}\label{fig:mono_exponential}}
\subfigure[]{\includegraphics[width=0.45\textwidth]{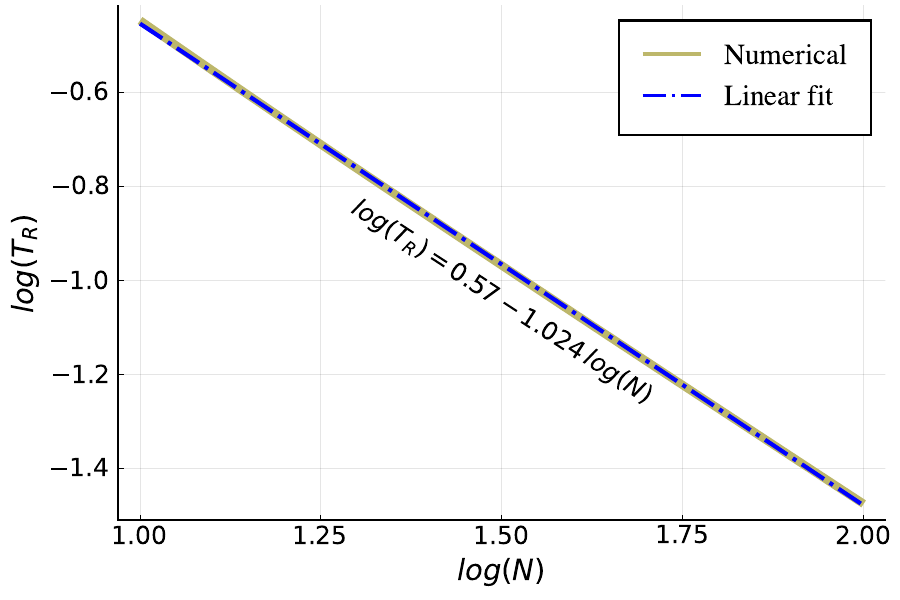}\label{fig:burst}}
\caption{The plot of von Neumann entropy as a function of $N$. Here $N=10$. For $f_{ab}=0$, as the system thermalizes, entropy behaves like an extensive quantity. On the other hand, for $f_{ab}=1$, the system skips thermalization, and the emerging conserved quantity leads to constant entropy for large $N$.  
We also plot the dynamics of $I = \langle J_+ J_- \rangle$ for $N=10$ with initial condition $\vert \psi (0) \rangle = \vert \uparrow_1 \uparrow_2 \dots \uparrow_N \rangle$.  
Forar $f_{ab}=0$, the dynamics are mono-exponential. For $f_{ab}=1$, a short-lived, intense radiation burst appears in an intermediate time-scale before reaching steady state.  
The maximum intensity plotted as a function of $N$ (from 10 to 100) shows a quadratic scaling: $\log I_{\max} = -0.618 + 1.9511 \log N$.  
The decay time $T_R$ shows inverse scaling: $\log T_R = -0.57 -1.024 \log N$.This short-lived, intense radiation is characteristic of Dicke superradiance.}
\label{fig:dicke}
\end{figure*}

For $
f_{ab}=0 $, the atoms are evolved through their individual dynamics,  hence the system reaches the Gibbs state at equilibrium. In this case, the steady state configuration is given by,
\begin{eqnarray}
    \rho_s^{eq} &=& e^{-\sum\limits_{i=1}^N \frac{\pi}{ \alpha} \hs^{\circ,N}}/\mathcal{Z}
\end{eqnarray}
 We note that $S$ is an extensive quantity for the system in a Gibbs state. Hence, for a fixed $\alpha$, using the above formula for $S$, one can show that $S \propto N$, which is consistent with the plot shown in Fig. \ref{fig:entropy}. 

The situation is entirely different for the $f_{ab} =1$ case. The corresponding dynamical equation for the $N$ spin case is given by,
\begin{eqnarray}
\frac{d \rho_s^N}{d\tau}&=& -\frac{i}{2} \sum \limits_{a,b=1}^{N} \sum \limits_{j,k=1}^3 \Big[\mathcal{S}_{jk}^{ab} \sigma _a ^j \sigma _b ^k~,\,\rho_s^N \Big] \nonumber\\
&&+ \sum \limits_{a,b=1}^{N} \sum \limits_{j,k=1}^3 \gamma^{ab}_{jk}\Big(\sigma _b ^k\rho_s \sigma _a ^j  
-\frac{1}{2}\{ \sigma _a ^j \sigma _b ^k ,\rho_s^N \} \Big)~ 
\label{eq:4a}
\end{eqnarray}
There exists an exchange symmetry in such a case (i.e., exchange of any $a$th and $b$th atom, the Liouvillian remains unchanged). Such symmetry leads to $N(N-1)/2$ numbers of conserved quantities for $f_{ab} =1$ (e.g., the corresponding operator is $\vec{\sigma}_a.\vec{\sigma}_b$), which shows the notion of integrability in the system \cite{Saha_2024}. 
The Liouvillian is invariant under the following unitary transformation, $\hat{U} \hat{\mathcal{L}} \hat{U}^\dagger = \hat{\mathcal{L}} $. Here $\hat{U} = e^{- i \hat{O}\phi }$, $\phi$ is a real parameter, and $\hat{O} = O\otimes \mathbb{I} - \mathbb{I} \otimes O^T$, with $O = \sum_{a,b=1}^N \vec{\sigma}_a.\vec{\sigma}_b$. The number of zero eigenvalues of $\hat{\mathcal{L}}$ increases exponentially with the system size. 
To explore the dynamics of the system, we use the collective basis for analysis. The collective operator is defined as, $\vec{J} = \sum_{i=1}^N \vec{\sigma}_i$, the correspond eigen-basis is defined as $\vert j m \rangle$. Here, $j = 1/2:N/2$, and $m = -j:j$. The number of independent blocks is $N/2 +1$ if N is even, else $(N+1)/2$. Similar to the two-spin case, the dynamics is confined to a particular block. We also note that the total population of each block depends on the initial configuration of the system $\rho_s(0)$. The separate irreversible dynamics of individual blocks show the notion of the Generalized Gibbs ensemble (GGE) \cite{Halati2022}. The dynamical equation of the individual block is given by,
\begin{eqnarray}
\frac{d \rho_s^J}{d\tau} &=&   \gamma_{-}\Big(J_-\rho_s J_+  
-\frac{1}{2}\{ J_+ J_- ,\rho_s \} \Big)  \nonumber \\
&&+  \gamma_{+}\Big(J_+\rho_s J_-  
-\frac{1}{2}\{ J_- J_+ ,\rho_s \} \Big)
\end{eqnarray}
$\rho_s^J$ is the density matrix corresponding to the $J$th block. We also neglect the effect of the Lamb shift in this case.
If the dynamics are initially confined in the principal block (i.e., $J = N/2$), then the corresponding steady state is given by,
\begin{eqnarray}
    \rho_s^{J, eq} &=& e^{- \frac{\pi}{\alpha} \omega_\circ J_z}/\mathcal{Z}_J \nonumber\\
    \mathcal{Z}_J &=& \tr( e^{- \frac{\pi}{\alpha} \omega_\circ J_z}) = \frac{\sinh{\frac{(N+1) \pi \omega_\circ }{2 \alpha}}}{\sinh{\frac{ \pi \omega_\circ }{2 \alpha}}}
\end{eqnarray}
Using the previous formula of $S$, we note that for a large $N$ limit, $S$ is independent of $N$, which is explicitly shown in Fig. \ref{fig:entropy}.
 Next, we focus on the dynamics of the system in the intermediate regime for both values of $f_{ab}$. 
\subsection{Emergence of Dicke superradiance}
\begin{figure}[!htbp]
    \includegraphics[width=\linewidth]{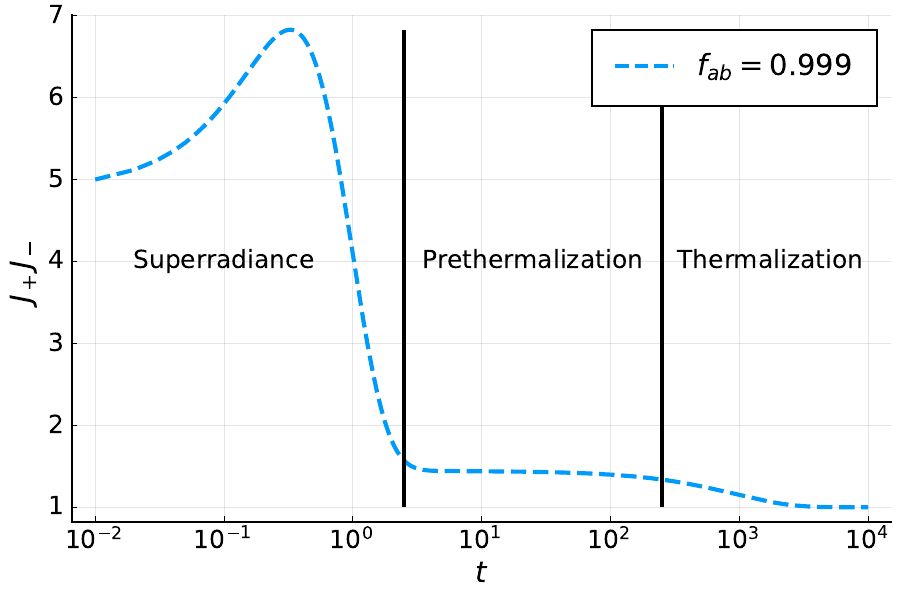}
    \caption{The plot of $\langle J_+J_- (t) \rangle$ vs $t$ for five atoms system with $f_{ab} = 0.999$. The initial state is $\vert \uparrow_1 \uparrow_2..\uparrow_5 \rangle$. The dynamics can be divided into three regions.  At a very short time scale, a Dicke superradiance-type radiative burst is observed. In the intermediate time-scale, it reaches a prethermal quasi-steady state. Finally, the prethermal generalized Gibbs state further decays to a thermal Gibbs state. }
    \label{fig:final}
\end{figure}

We find that the main difference between $f_{ab} \to 1$ and $f_{ab}<1$ is the presence of collective dissipation at $f_{ab} \to 1$. Hence, we study the emerging collective features that help to distinguish it from Unruh thermalization. Recently, multi-atom superradiation in a cylindrical cavity via vacuum fluctuation was analyzed in the context of detecting the Unruh effect \cite{Zheng2025}. We have a similar kind of setup. However, the major difference in our work is the presence of linear acceleration instead of circular motion. We study the time evolution of the radiative intensity operator, which is defined as $I = J_+J_-$. The initial state is chosen as $\vert \psi (0) \rangle = \vert \uparrow_1 \uparrow_2..  \uparrow_N \rangle$, or $\vert \psi (0) \rangle = \vert N/2, N/2 \rangle$  For $f_{ab}<1$, $\langle J_+J_- (t) \rangle$ shows an exponential decay with time (Fig. \ref{fig:dicke_entropy}). On the other hand, for $f_{ab} \to 1$, $\langle J_+J_- \rangle$ shows the feature of an intense and short-time radiative burst, which is non-mono-exponential and has several similarities with Dicke super-radiance (Fig. \ref{fig:dicke_entropy}). We plot the maximum intensity by varying the number of atoms, and we find that $I_{\max}  \propto N^2$ in Fig. \ref{fig:mono_exponential}. The lifetime of the burst can be calculated from the eigenvalue analysis of the block-diagonalized Liouvillian. The asymptotic decay rate (ADR) $\Delta$ is defined as the difference between the real value of the zero eigenvalue and the nearest negative eigenvalue \cite{Buca_2012}. The inverse of ADR is proportional to the decay time. Here, we plot  $1/\Delta$  as a function of $N$. We show that $\text{decay time} \propto 1/N$ (Fig. \ref{fig:burst}). Hence, the radiative burst is short-lived and intense and similar to Dicke superradiance. On the other hand, the decay time is constant for the Unruh thermalization case, as the individual atoms reach the steady state separately at the same time. We also show the plot of $\langle J_+J_- (t) \rangle$ vs $t$ for a 5-atom system at $f_{ab} = 0.999$, to show exact dynamics before it thermalizes in Fig. \ref{fig:final}. For Unruh thermalization, the dynamics are always mono-exponential towards the Gibbs state. However, Unruh prethermalization at $f_{ab} \to 1$, the dynamics show an initial radiative burst before reaching the prethermal state. Finally, the prethermal state shows a mono-exponential decay to the thermal state. We mostly focus on the collective behavior of the system in this section. Such non-equilibrium phenomena also have acceleration dependence, which we skip in our analysis as it is quite similar to the generalized Gibbs state. As such, if the prethermal magnetization value increases, then the maximum peak will also increase.

\section{Comparison with the experimental observations}
\label{sec-v}
The correspondence between the Unruh temperature and the relativistic acceleration is given by, $\kappa T = \hbar \alpha/2 \pi c$ \cite{takagi1986vacuum}. Here $\kappa$ is the Boltzmann constant, $c$ is the velocity of light, and $\hbar$ is the Planck constant. The above relation shows that for producing $1K$ temperature, the limiting value of relativistic acceleration $\alpha = 2.4 \times 10^{20}\,  m/s^2$.  Previous experiments were done in a circular ring using the presence of centrifugal acceleration. Using a 3.1 km accelerator ring of SPEAR at Stanford, the ultra-relativistic electron can be rotated at the value of $\alpha = 2.9 \times 10^{23}\, m/s^2$. Such experiments can be used to produce a range of temperatures from ten kelvin to a thousand Kelvin \cite{Bell1983}. 
\begin{figure}[!htbp]
    
    \includegraphics[width=\linewidth]{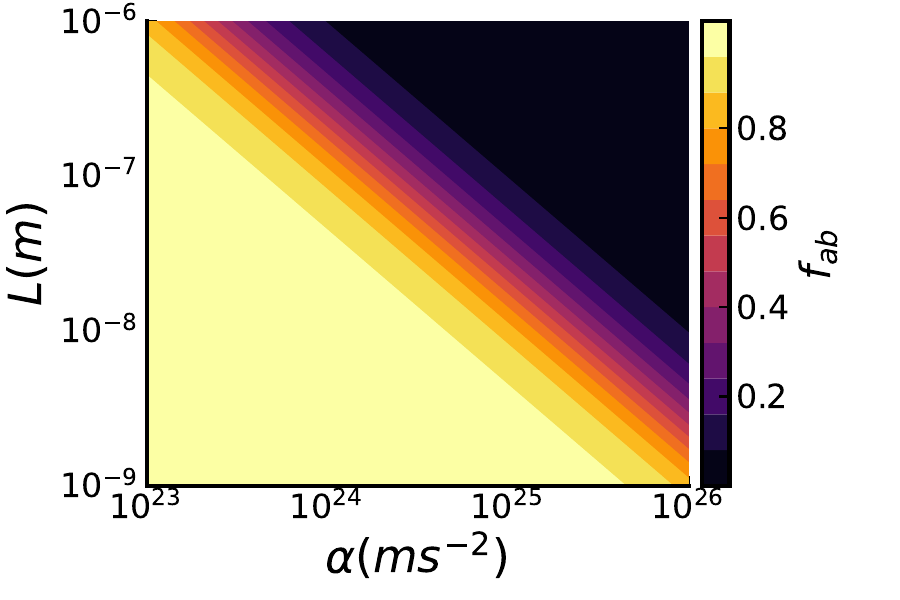}
    \caption{The contour plot show the value of $f_{ab}$ for changing $L$ and $\alpha$. We choose $\omega_\circ = 10^{15}$ Hz. The yellow region corresponds to $f_{ab} \to 1$, where the system shows the notion of prethermalization.}
    \label{fig:fab}
\end{figure}
For synchrotron radiation, the ultra-relativistic electron needs to rotate in a magnetic field at a critical frequency of $10^{15}$ Hz; hence, we choose $\omega_\circ = 10^{15}$ Hz for our calculation. We are interested in the values of the minimum inter-atomic distance for exhibiting such phenomena. Our calculation shows that for varying the values of $\alpha$ between  $10^{23}$ to $10^{26}\, m/s^2$, we get $f_{ab} \to 1$ when the average distance between the atoms is between $\mu m$ to $nm$ range (the yellow region in the figure). Previous experiments show that the minimum distance between the conducting plane for observing the Casimir-Polder interaction is $0.5 - 3.0 \, \mu m$ \cite{Bressi2002}. Hence, using similar atomic separation and the ultra-relativistic centrifugal acceleration used in synchrotron radiation, one can observe that the system will skip the Unruh thermalization and reach a prethermal state. We note that the system generally reaches the thermal state at a time scale proportional to $ T_{\rm th} = 1/\{(\gamma_+ + \gamma_-)(1-f_{ab})\}$ (shown in Fig. \ref{fig:enter-label} ) and reaches the prethermal state at a time scale proportional to $T_{\rm pre} = 1/(\gamma_+ + \gamma_-)$. We define fractional prethermal lifetime $T_{\rm pre}^{\rm frac}$ as,
\begin{eqnarray}
    T_{\rm pre}^{\rm frac} = \frac{T_{\rm th} -T_{\rm pre}}{T_{\rm th}} 
    \propto  f_{ab}
\end{eqnarray}
We note that $T_{\rm pre}^{\rm frac} $ depends only on $f_{ab}$, hence such a quantity can be identified using the following experimental values shown in Fig. \ref{fig:fab}.
\section{Discussions and Outlook}
\label{sec-vi}

We have revisited the well-known thermalization theorem originally proposed by Unruh and Davies in the context of uniformly accelerating many-body systems. In the case of RCPI, for a small distance $\alpha L \ll 1$, a local inertial frame description is valid, so the response is thermal. However, for the opposite limit $\alpha L \gg 1$, such a description is invalid, and a breakdown of thermal response is reported \cite{Marino2014}. An interesting observation has also been made for the steady-state configuration of such systems \cite{Benatti_2002}. The scalar field induced entanglement can be possible in the regime $\alpha L \ll 1$. A common question that arises is how entanglement can be generated in such regime. We understand that such an answer was still missing in the literature.. Hence, in this manuscript, we provide a brief analysis of system dynamics in this regime. 

The consequence of the thermalization is the following: the initial memory of the system vanishes, and the system reaches the Gibbs state. Surprisingly, in the case of the Unruh effect, similar things can be observed, where the role of temperature is played by acceleration. In condensed matter physics, for both closed and open quantum systems, the integrability shows a sharp distinction from such thermalization phenomena \cite{Saha_2024}. It is expected that there should also be a one-to-one correspondence in the case of the Unruh effect. We show that for the $\alpha L \ll 1$ case,  the system becomes nearly-integrable as there exists an extensive number of quasi-conserved quantities. Such phenomena are quite similar to prethermalization in open quantum systems. 

Unruh thermalization is related to the dissipation of the system coupled to a local bath. As a result, the atoms individually reach the Gibbs state at equilibrium. The questions remain about the influence of collective dissipation, which has not been investigated previously within the framework of Unruh thermalization.  Previous works were confined to the two-atom case, where the entanglement generation is extensively studied.  Here we provide the generalized picture of the two-atom case \cite{Benatti_2002}. Our result shows that for the $\alpha L \ll 1$ case, the collective dissipation dominates over the local dissipation. The long-time regime shows the notion of the Unruh effect. However, in the intermediate time-scale, the collective dissipation results in the integrability as the dynamics is confined in a symmetric subspace. Similarly, other collective feature, i.e., Dicke superradiance, is also observed before the systems reach the prethermal state.
Our recent work on linear acceleration can also be applied to  the the ultra-relativistic particles in circular motion 
 \cite{Bell1983}, for the future experimental verification of Unruh prethermalization.

In this manuscript, we are not claiming any modification of the Unruh thermalization theorem. However, such a theorem doesn't represent the whole story for the non-interacting many-atom systems, as in such cases, a cascaded dynamics is observed, as the system reaches a long-lived prethermal state in the intermediate time-scale before it thermalizes, which has not been previously explored.

We recognize several compelling areas where our present work can be explored further. Given that most experimental setups emphasize circular acceleration rather than linear, a critical investigation into the lifetime of the prethermal state under centrifugal acceleration presents a significant opportunity as our work predicts that using the relativistic particles in circular ring also shows prethermalization.  Moreover, enhancing bath-induced entanglement through the memory effects of non-Markovian baths offers several benefits in quantum information. 
Additionally,  the influence of non-Markovian vacuum fluctuations on Unruh prethermalization is also an interesting area to gain more insights into these complex interactions.

\section{Conclusion}
As a summary, we discuss the correspondence between the open quantum
system and uniformly accelerated non-interacting atoms. For a single 
atom, it is always reach the thermal state. On the other hand, for a many-spin case, the system may skip the thermalization, as in $\alpha L \ll 1$, 
the dynamics is constrained by the extensive number of quasi-
conserved quantities. Hence, such constrained dynamics lead 
to Unruh prethermalization. We also show that in that regime, the 
collective dissipation dominates. As such, the system shows a Dicke superradiance-type collective emission process before reaching the prethermal state. Finally, we 
also predict the values of $\alpha,\, L,\, \omega_\circ$ in 
connection with the experiments using accelerating electrons in a circular ring.
\label{sec-vii}

\section*{Acknowledgments}
 SS gratefully acknowledges the University Grants Commission for a PhD fellowship (Student ID: MAY 2018- 528071). A. C. acknowledges support from the University Grants Commission with a PhD fellowship (reference ID: 522160) and the São Paulo Research Foundation (FAPESP) (Process No. 2024/22111-3) for carrying out the research.

\bibliographystyle{apsrev4-1}
\bibliography{ref2}
\end{document}